\documentstyle[aps,multicol,prl]{revtex}

\title{Maximum entropy approach to
power-law distributions in coupled dynamic-stochastic systems.} 
\author{E.~V.~Vakarin, J.~P.~Badiali} 
\address{ UMR 7575 LECA ENSCP-UPMC-CNRS, 11 
rue P. et M. Curie, 75231 Cedex 05, Paris, France }

\begin{document}
\maketitle
\begin{abstract}
Statistical properties of coupled dynamic-stochastic systems are 
studied within a combination of the maximum
information principle and the superstatistical approach. The conditions
at which the Shannon entropy functional leads to a power-law statistics
are investigated. It is demonstrated that, from a quite general point of
view, the power-law dependencies may appear as a consequence of  "global"
constraints restricting both the dynamic phase space and the stochastic
fluctuations. As a result, at sufficiently long observation times the
dynamic counterpart is driven into a non-equilibrium steady state whose
deviation from the usual exponential statistics  is given by the distance  
from the conventional equilibrium. 
\end{abstract}


\begin{multicols}{2}

\section{Introduction}
Power-law distributions are quite common for complex systems of different 
nature\cite{Newman}. Several theoretical schemes have been developed in 
order to understand this behavior. The concepts of the self-organized 
criticality\cite{Bak} and highly optimized tolerance\cite{Carlson}  
have been extensively illustrated within various kinetic
(sandpile, slider-block, forest fire, etc.) models\cite{Carlson,SOC}, which 
exhibit a triggering between different regimes, accompanied by power-law 
dependencies - avalanches. More recent studies reported on similar
phenomena in inelastic dissipative gases\cite{cascade}, stochastic processes
with multiplicative noise\cite{noise}, clustering models\cite{Powerdif},
and condensation in porous media\cite{avalanches}.
Observations of the power-low features in a variety of much more complex
(physical, biological, social, etc.) 
systems\cite{SOC,Powerdif} led to further searches for a generic  
mechanism, responsible for such a behavior independently of a system 
microscopic specificities or model approximations.

A remarkable step was to find\cite{Wilk} that a power-law could 
appear from the usual exponential with fluctuating parameters. This is a  
basic idea  of the superstatistical approach\cite{Beck,Beck1,scales} 
explaining the power-law statistics in a system as a result of fluctuations 
in its surrounding (background). The latter can be modeled as a stochastic
\cite{Wilk,Beck,Beck1,scales,mass} or a dynamic 
(Hamiltonian)\cite{Extended,separation} process. 
Nevertheless the information on
the source of the background fluctuations is not always available.
Moreover, in many cases (e.g. optimal control\cite{Touchette} or insertion
into flexible environments\cite{info}) the two counterparts are 
strongly coupled and the background statistics is (or should be) conditional 
to the system properties. In that case the background evolution cannot be
modelled as a process (e.g. fluctuating temperature\cite{Beck} or 
mass\cite{mass}) independent of the system state.  
Lacking any detailed information on the coupling, one has to 
resort to the inference methods. 

The maximum entropy inference methods\cite{Jaynes} have been
developed in this context. These are based on parameterized information
entropy measures (Tsallis\cite{Tsallis} or Renyi\cite{Jizba,Bashkirov})
which have the Shannon form as a limit. Despite of various successful 
applications, this approach has generated several controversial points
\cite{Critique} related to the meaning of the entropic index,   the 
non-additivity effects, spurious correlations and the biased averaging
\cite{Vives}. This stimulated further steps\cite{Sattin,Dover} towards 
deriving non-exponential distributions by maximizing the Shannon entropy
under suitable constraints.

Therefore, it seems promising to develop a scheme, capable of
merging the advantages of the maximum entropy and the superstatistical
approaches. More specifically, our main goal is to find out the conditions
at which the Shannon entropic form leads to the background statistics,
coherent with the one introduced within the superstatistical approach.
This would allow us to clarify the 
mechanism of the corresponding power-law dependencies  and to 
understand the essence of the non-additivity effects.
 
\section{Coupled systems}
 
With this purpose we consider a many-body {\it dynamic} 
(deterministic) system in contact with a fluctuating background. 
In contrast to coupled dynamic systems\cite{PRL}, in our case the two
subsystems are of different origin and thus require different 
levels of description\cite{CEJP}. 
The background (the {\it 
stochastic system}) is considered to be a source 
of some relevant 
quantity, $\beta$ (e. g. pore size, temperature, etc), 
which fluctuates according to a probability distribution 
$f(\beta)$. 
Quite often the different nature of the two subsystems leads to a 
well-defined separation of the relaxation time scales\cite{scales,mass}, 
typical for self-organizing or glassy systems. Namely, the system relaxation
is supposed to be much faster than the background fluctuations. 
Such that for any background 
state the system can reach an equilibrium (or stationary) state with a 
conditional thermodynamic function $\theta(\rho|\beta)$. In principle, this 
could be any function, suitable for an adequate description of the 
system internal order and linking 
the relevant extensive parameters $\rho$ to the 
intensive ones $\beta$. In what follows the function $\theta(\rho|\beta)$
is assumed to be known (from exact results or relevant approximations).  
Then an observable can be represented as an
average over the background fluctuations 
\begin{equation}
\label{constraint}
T(\rho)={\overline {\theta(\rho|\beta)}}=
\int d\beta f(\beta)\theta(\rho|\beta) 
\end{equation}
where the overbar denotes the corresponding averaging. Note that the
time scales separation is essential, otherwise the "quenched" average
in (\ref{constraint}) does not make sense. 

If the background does not undergo some internal stochastic
process independently of the system, then $f(\beta)$ is {\it a priori}
unknown. It should be determined from the information on $T(\rho)$. 
This problem is typical for characterizing the heterogeneous media
through indirect (e.g. adsorption) probes, where $T(\rho)$ is a measurement
result. 
On the other hand, in many
applications (e.g. protecting storage, adaptive learning and 
control\cite{Touchette,Dover}) it is desirable to design the background 
in the way leading to a well-defined behavior $T(\rho)$. In this case
$T(\rho)$ should be considered as a cost function.
\section{Maximum entropy approach}
Therefore, one deals with an inverse problem of extracting $f(\beta)$
from $T(\rho)$. This can be done within the maximum-entropy
inference scheme proposed by Jaynes\cite{Jaynes}.
Our uncertainty on the background state can be estimated by an information 
entropy, which is taken in the Shannon form. 
\begin{equation}
\label{Shannon}
H=-\int d\beta f(\beta) \ln [f(\beta)]
\end{equation}
Maximizing $H$ under the constraint (\ref{constraint}) and requiring
the normalization for $f(\beta)$ we get the following 
conditional distribution $f(\beta)=f(\beta|\rho)$ \cite{info}
\begin{equation}
\label{distribution}
f(\beta|\rho)=\frac{e^{-\kappa 
\theta(\rho|\beta)}}{Z}; \qquad
Z=\int d\beta e^{-\kappa\theta(\rho|\beta)}
\end{equation}
where the Lagrange multiplier $\kappa$ should be found 
from the constraint (\ref{constraint}), that is equivalent to solving
$$
T(\rho)=-\frac{\partial}{\partial \kappa} \ln Z
$$ 
Plugging the distribution (\ref{distribution}) back to (\ref{Shannon})
we obtain the amount of information (on the background) one can 
get by driving (e.g. through varying $\rho$) the dynamic subsystem 
\begin{equation}
H(\rho)=\kappa T(\rho)+\ln Z
\end{equation}
In particular, the information rate takes a remarkably simple form
\begin{equation}
\frac{\partial H(\rho)}{\partial \rho}=
\kappa\left[
\frac{\partial T(\rho)}{\partial \rho}-
{\overline 
\frac{\partial \theta(\rho|\beta)}{\partial \rho}
}
\right]
\end{equation}
It is clear that the scheme gives a solution for $f(\beta)$ which
is free from adjustable parameters, providing an explicit link between
the data (or cost function) and the conditional theoretical estimation.
Therefore, the sensitivity to the "kernel" variation
$\theta(\rho|\beta)$ and to scattered data $T(\rho)$ can be easily 
controlled. On the other hand, if $f(\beta)$ is already known, then 
our results can be used to estimate the quality of the model
$\theta(\rho|\beta)$ through its matching to the available data
$T(\rho)$.
\section{Constraints}
Despite of the apparently exponential form (\ref{distribution}) the actual
behavior of $f(\beta)$ depends on the nature of the constraint imposed and 
on a form of the constrained function $\theta(\rho|\beta)$. Such an ambiguity
should not be considered as a shortcoming of the theory. This is a 
consequence of the fact that we are working under conditions of incomplete 
information. Then, according to the Baesian interpretation,  a 
probability should be considered as a measure of our ignorance rather than 
an objective property. Nevertheless, our freedom in choosing 
the constraints is restricted by any prior information, coming through
independent tests.     
On the other hand, there are constraints which are "naturally" imposed 
either as design principles\cite{Network} or as experimental conditions. In 
what follows we discuss two relevant examples.  
\subsection{Entropy constraint}
As is mentioned earlier, in many applications it is desirable
to constraint the system internal order with the purpose of
meeting some survival or functionality objectives. In this case
it is natural to restrict the phase space by constraining the thermodynamic 
entropy $S(\rho|\beta)$. This idea was shortly discussed in a slightly
different context \cite{info,Dover}. Therefore, we set  
$\theta(\rho|\beta)=S(\rho|\beta)$, $T(\rho)=\Sigma(\rho)$. Then the 
distribution (\ref{distribution}) closely resembles the Einstein fluctuation
formula. Note however that $f(\beta)$ describes the background fluctuations
and for $\kappa=1$ it becomes identical to the distribution of the system
fluctuations. In particular, for small fluctuations around an equilibrium
state $(\rho,\beta^*)$ we may expand
\begin{equation}
\label{quadr}
S(\rho|\beta)=S(\rho|\beta^*)-\frac{1}{2\chi(\rho,\beta^*)}
(\beta-\beta^*)^2
\end{equation} 
where $\chi(\rho,\beta^*)=\langle(\beta-\beta^*)^2 \rangle $ is the
mean-square fluctuation in the system when the background state
is fixed at $\beta=\beta^*$.  
In this approximation the distribution (\ref{distribution}) becomes
gaussian and $\kappa$ can be determined combining (\ref{quadr}) and 
(\ref{constraint}). Finally we arrive at
\begin{equation}
\label{fluct}
{\overline{(\beta-\beta^*)^2}}=\langle(\beta-\beta^*)^2 \rangle
\left[
S(\rho|\beta^*)-\Sigma(\rho)
\right]
\end{equation}
Note that $S(\rho|\beta^*)$ is the system equilibrium entropy. Consequently
$S(\rho|\beta^*)-\Sigma(\rho)\ge 0$ and ${\overline{(\beta-\beta^*)^2}}\ge 
0$ as it should be. Therefore, in order to ensure a given response, 
$\Sigma(\rho)$, the background 
 should fluctuate coherently with the system fluctuations  
and with the distance from the equilibrium state. As we will see below,
for large fluctuations this tendency also takes place.  
A quite similar trend  
has been reported\cite{Vives} for fluctuations in the Tsallis statistics.
In the limit of $S(\rho|\beta^*)=\Sigma(\rho)$ we return to the standard
equilibrium without fluctuations in the background:
$f(\beta)=\delta(\beta-\beta^*)$. The system  fluctuates according
to its response function $\chi(\rho,\beta^*)$.   

In order to study large background fluctuations and 
the system statistics at different time scales
we have to introduce an explicit form for 
$S(\rho|\beta)$. With this purpose we consider an exactly 
solvable toy model -- the ideal gas in contact with a reservoir of 
fluctuating temperature\cite{Beck,Beck1}. This choice is motivated
by our goal to extract the most general and essential features, independent 
of approximations or the system correlations. Therefore, we deal with
the entropy per particle $S(\rho|\beta)=5/2-\ln(\rho \Lambda^3)$. Here
$\rho$ is the number density, $\beta=1/kT$ is the inverse temperature
 and $\Lambda$ is the thermal de Broglie 
length. Introducing irrelevant scaling constants (making $\rho$ and $\beta$ 
dimensionless), $S(\rho|\beta)$ can be reduced to 
\begin{equation}
S(\rho|\beta)=const-\ln(\rho)-\frac{3}{2}\ln(\beta)
\end{equation}
Constraining the average temperature  
\begin{equation}
\beta_0=\int d\beta f(\beta)  \beta
\end{equation}
and the entropy 
\begin{equation}
\label{th-entropy}
\Sigma(\rho)=\int d\beta f(\beta)S(\rho|\beta) 
\end{equation}
through the inference procedure discussed above, 
we obtain the following distribution 
\begin{equation}
f(\beta)=\frac{\beta^{\kappa} e^{-\beta/\beta(\kappa)}}
{[\beta(\kappa)]^{\kappa+1}\Gamma(\kappa+1)};
\quad
\beta(\kappa)=\frac{\beta_0}{\kappa+1}
\end{equation}
which is precisely the $\Gamma$-distribution considered in the 
superstatistical approach\cite{Beck,Beck1}, where the exponent
$\kappa$ is related to the noise intensity.   
In our case the Lagrange multiplier $\kappa$ should be determined from the 
entropy constraint (\ref{th-entropy})
\begin{equation}
\Sigma(\rho)=S(\rho|\beta_0)+\ln(\kappa+1)-\Psi(\kappa)-\frac{1}{\kappa}
\end{equation} 
where $\Psi(\kappa)=d\ln \Gamma(\kappa)/d\kappa$. Thus, the exponent 
$\kappa$ is determined by the distance $\Sigma(\rho)-S(\rho|\beta_0)$
from the equilibrium state ($\rho,\beta_0$). In particular, 
for large $\kappa$ we have found
$$
\Sigma(\rho)=S(\rho|\beta_0)+1/(2\kappa)
$$
Therefore, $\kappa$ is related to the deviation from the standard
equilibrium, such that $f(\beta) \to \delta(\beta-\beta_0)$
and $\Sigma(\rho)=S(\rho|\beta_0)$ as $\kappa \to \infty$. 

At the short-time scale (of the order of the system relaxation time)
the conditional  energy distribution ($E= p^2/2m$, at a given temperature 
$\beta$ ) is Gibbsian 
\begin{equation}
f(E|\beta)=\frac{e^{-\beta E}}{\int dE e^{-\beta E}}
\end{equation}
The long-time behavior of the dynamic system 
can be represented as a superposition of its short-time statistics and the 
background fluctuations -- the superstatistical approach\cite{Beck,Beck1}. 
In this spirit the long-time energy distribution can be found by averaging 
over the temperature fluctuations
\begin{equation}
f(E)=\int d\beta f(\beta)f(E|\beta)=
\beta_0 \left[
1-\frac{\beta_0 E}{1-q}
\right]^{-q} 
\end{equation}
where $q=\kappa+2$. 
For $E= p^2/2m$ we recover the power-law velocity
distribution found\cite{Beck,Beck1} in  the superstatisical approach.
Quite similar effects have recently been predicted to occur in
driven dissipative inelastic gases\cite{cascade} and in driven stochastic
systems with multiplicative noise\cite{noise} or fluctuating mass\cite{mass}
In a different context similar power laws where found, applying the 
maximum-entropy inference to parameterized entropies (Tsallis\cite{Tsallis}, 
and Renyi\cite{Bashkirov}).  But the meaning of the entropic parameter is 
not always clear, while in our case $q$ is directly related to the 
constraint imposed.

Thus, because of the constraint, imposed on 
the internal order at longer times the system develops avalanches (or energy 
cascades) at any finite $\kappa$.
The avalanche size is characterized by $1/\kappa$. Therefore, maintaining
the distance from the equilibrium, one can control the size of the
rare "catastrophic" events. This can be organized in different ways, such as
by powerful energy injections at large velocity scales\cite{cascade}, or
through an interplay of additive and multiplicative noises\cite{noise}.

\subsection{Activity constraint}
Other constraints are realized as driving conditions. For instance, 
adsorption into porous media\cite{avalanches} is driven by a difference
between the chemical potential in the fluid bulk, $\mu_b$ and that
inside the matrix $M(\beta,\rho)$. Adsorption equilibrium corresponds to
$M(\beta,\rho)=\mu_b$. Nevertheless, recent analysis\cite{avalanches} of 
experimental results reveals that the true equilibrium seems to be
hardly reachable because of very long equilibration times and well-developed
metastability. At certain conditions this leads to the hysteretic behavior
accompanied by multiple metastable states of the fluid.  
On the other hand, a complicated matrix topography makes one
to resort to a statistical description (e.g. pore sizes, site energies, 
etc.). In this context the matrix can be  
considered as medium inducing a distribution $f(\rho_1...\rho_N)$ of 
metastable states with local 
fluid densities $\rho_i$ in different spatial domains $i=1..N$. For 
simplicity the domains are assumed to be uncorrelated: 
$f(\rho_1...\rho_N)=\prod_i f(\rho_i)$. If the temperature $\beta$ is fixed, 
then the fluid state in each domain is given by a local isotherm 
$\mu(\beta|\rho_i)$. The overall isotherm is an average over the domains
\begin{equation}
\label{mu}
M(\beta,\rho)=\int d\rho_i f(\rho_i)\mu(\beta|\rho_i)
\end{equation} 
where $\rho$ is the average fluid density given by 
\begin{equation}
\label{rho}
\rho= \int d\rho_i f(\rho_i) \rho_i
\end{equation} 
The local isotherm is chosen in the ideal gas form
$\beta\mu(\beta|\rho_i)=\ln(\rho_i\Lambda^3)$.
Maximizing the Shannon entropy under constraints (\ref{mu}), (\ref{rho})
we obtain the following local density distribution 
\begin{equation}
\label{distribution1}
f(\rho_i)=\frac{[\rho_i/\rho(\lambda)]^{\lambda} e^{-\rho_i/\rho(\lambda)}}
{\rho(\lambda)\Gamma(1+\lambda)};
\quad
\rho(\lambda)=\frac{\rho}{1+\lambda}
\end{equation}
which is (for $-1<\lambda<0$) of the form taken in \cite{avalanches} as a 
fitting function for the description of the collective condensation events 
(avalanches). The Lagrange multiplier $\lambda$ should be determined from 
the constraint (\ref{mu}) 
\begin{equation}
M(\beta,\rho)=\mu(\beta|\rho_i=\rho)-\ln(\lambda+1)+\Psi(\lambda)+
\frac{1}{\lambda}
\end{equation}
Therefore, the exponent $\lambda$ is again related to the distance 
$M(\beta,\rho)-\mu(\beta|\rho_i=\rho)$ from the conventional equilibrium.
The latter appears as a limit of $\lambda \to \infty$, leading to 
 $f(\rho_i)=\delta(\rho_i-\rho)$ when 
$M(\beta,\rho)=\mu(\beta|\rho_i=\rho)$.

It should be noted that the driving procedure is extremely important.
If, for instance, the chemical potential (pressure) is free to relax 
according to a controlled particles injection, then the constraint 
(\ref{mu}) should be removed ($\lambda=0$) and the system follows a 
different path with a purely exponential distribution. This agrees with the 
conclusion made in \cite{avalanches}.
Moreover, it can be easily shown that a dependence on the driving path
is a generic feature of the coupled systems considered here. For this 
purpose let us consider an adsorption of noninteracting species into a 
network with fluctuating site binding energy $\epsilon$, distributed 
according to some probability density $f(\epsilon)$. If the process is 
driven by increments in the chemical potential $\mu$, then the conditional 
grand potential is 
\begin{equation}
\Omega(\mu|\epsilon)=-\ln
\left(
1+e^{\mu+\epsilon}
\right)
\end{equation} 
The adsorption isotherm (coverage $T$ vs $\mu$) is given by
\begin{equation}
\label{T}
T(\mu)=-{\overline{\frac{\partial \Omega(\mu|\epsilon)}{\partial \mu}}}=
{\overline{\left(\frac{e^{\mu+\epsilon}}{1+e^{\mu+\epsilon}}\right)}} 
\end{equation} 
On the other hand, if the coverage $\theta$ is maintained by controlled
injections, then the conditional free energy is
\begin{equation}
F(\theta|\epsilon)=-\epsilon\theta +\theta\ln\theta+(1-\theta)\ln(1-\theta)
\end{equation}
This allows to calculate the chemical potential
\begin{equation}
\mu={\overline{\frac{\partial F(\theta|\epsilon)}{\partial\theta}}}=
-{\overline{\epsilon}}+\ln{\frac{\theta}{1-\theta}} 
\end{equation} 
which can be formally solved with respect to $\theta$
\begin{equation}
\label{theta}
\theta=\frac{e^{\mu+{\overline{\epsilon}}}}
{1+e^{\mu+{\overline{\epsilon}}}}
\end{equation}
Therefore, the isotherm strongly depends on the driving conditions.
The main reason is a non-zero distribution width (fluctuations) in the
stochastic counterpart. This can be demonstrated by expanding (\ref{T}), 
(\ref{theta}) in terms of $\epsilon$ and analyzing the difference  
\begin{equation}
T-\theta=\sum_{n=2}^{\infty}a_n(\mu)
\left[{\overline{\epsilon^n}}-{\overline{\epsilon}}^n\right]
\end{equation}
which vanishes only if the distribution is $\delta$-like (no fluctuations).
Note that our conclusion is quite general. It does not depend on a 
shape of the distribution $f(\epsilon)$ and on the adsorbate dynamics.
\section{Conclusion}
We have found that the major ingredients, relevant to the power-law 
distributions in composite systems are (i) the widely separated times scales,
(ii) non-vanishing background fluctuations. 
(iii) a constraint, imposed on the overall system, holding the dynamic
counterpart in a stationary non-equilibrium state.

One might argue  that the $\Gamma$- and the power-law 
distributions result trivially from the logarithmic form of the constrained 
functions. It should be noted, in this respect, that the logarithmic shape 
of the thermal entropy is of quite general nature, as this follows from the 
famous Boltzmann relation $S=k\ln W$. The density distribution 
(\ref{distribution1}) is also not a specific feature of the ideal gas. Our 
results do not alter if the interparticule interactions are taken into 
account (e.g. as a long-range perturbation\cite{JCP}): 
 $$
\beta\mu'(\beta|\rho_i)=\ln(\rho_i\Lambda^3)-U\rho_i
$$
In that case the distribution (\ref{distribution1}) does not change and
the exponent $\lambda$ is determined by the distance from the equilibrium
state $\mu'(\beta|\rho_i=\rho)$. 

The main advantage of the approach developed here is that it avoids 
parameterized entropy measures and allows one to apply the superstatistical 
scheme to coupled systems in which the stochastic background does not
fluctuate independently of the dynamic counterpart.

In the context of our study the nonextensitivy in
systems with power-law distributions is a direct consequence
of the "global" nature of the constraint. This makes it impossible
to decompose the system into non-correlated parts. Moreover, for any
background distribution of non-zero width, the system thermodynamic
response strongly depends on a driving path.     

 
\end{multicols}
\end{document}